# Photodegradation and Thermal Effects in Violet Phosphorus

Mahdi Ghafariasl[1], Sarabpreet Singh[1], Sampath Gamage[1], Timothy Prusnick[2], Michael Snure[3] and Yohannes Abate[1*]

[1] Department of Physics and Astronomy, University of Georgia, Athens, GA 30602, USA

[2] KBR, Beavercreek OH 45433, USA

[3]Air Force Research Laboratory, Sensors Directorate, Wright Patterson Air Force Base, Ohio 45433, USA

**Abstract**

Violet phosphorus (VP) has garnered attention for its appealing physical properties and potential applications in optoelectronics. We present a comprehensive investigation of the photo-degradation and thermal effects of exfoliated VP on $SiO_2$ substrate. The degradation rate of VP was found to be strongly influenced by the excitation wavelength and light exposure duration. Light exposure to above bandgap ($\lambda > 532$ nm) leads to faster degradation, attributed to interactions with reactive oxygen species. Power-dependent photoluminescence (PL) measurements at low temperature (T=4 K) showed neutral exciton ($X^0$) and trion (T) intensities linearly increased with excitation power, while the energy difference between their peak energies decreased, indicating changes in the exciton energy gap due to degradation. At room temperature $X^0$ and T peaks were observed with higher $X^0$ spectral weight, indicating reduced thermal stability of T. As the temperature decreased to 4 K, both $X^0$ and T emissions intensified with blue-shifted peak positions. The $T/X^0$ spectral weight ratio increased from 0.28 at 300 K to 0.69 at 4 K, suggesting enhanced T formation due to reduced phonon scattering. Temperature-dependent Raman spectroscopy revealed the presence of VP up to 673K. By tracking the peak position of 9 Raman modes with temperature the linear first-order temperature coefficient were obtained and found to be linear for all modes up to 673 K. Our results provide a deeper understanding of VP's degradation behavior and implications for optoelectronic applications.

## Introduction

The ongoing evolution of the rich phase diagram of phosphorus allotropes, paired with the successful synthesis of these structures, has sparked significant interest and opened avenues for diverse fundamental research and applications[1-11]. Among these is violet phosphorus (VP), a layered elemental semiconductor with a bandgap tunable in the range 1.5 -2.5 eV[5, 12-14]. Compared to the other phosphorus allotropes, VP is the most stable making it a promising candidate for research and optoelectronic applications[9, 15-19]. Recently research has focused on various aspects of VP degradation under different environmental conditions. Factors such as substrate hydrophobicity and surface roughness can influence VP's stability by modifying van der Waals forces[20]. Comparisons between VP and Black phosphorus (BP) indicate that VP exhibits greater stability under ambient conditions in terms of surface chemical degradation[21]. The effect of oxygen and water in the degradation of VP has been studied [13] , needle-shaped VP have also been used as a catalysts[22]. Thermal degradation studies have explored the evolution of Raman modes, some of which exhibit non-linear temperature dependence[23].

In this work, we studied the photodegradation and thermal effects of exfoliated VP on a $SiO_2$/Si substrate using atomic force microscopy (AFM), Raman, and Photoluminescence (PL) spectroscopy across a range of temperatures and wavelengths. Our findings revealed a distinctive modulation in the Raman intensity of VP, shedding light on the oxidation mechanism through optical excitation. The degradation rate of VP was influenced by the excitation wavelength and exposure duration, with above bandgap excitation leading to faster degradation due to interactions with reactive oxygen species (ROS) that are generated by the incident laser. PL spectroscopy analysis indicated a gradual decline in the exciton population, suggesting reduced lifetime and alterations in formation and stability of excitons, impacting VP's quantum efficiency. Power-dependent PL measurements demonstrated linear intensity increases in neutral excitons and trions, while the energy difference between their peak energies decreased with increasing power, indicative of changes in the exciton energy gap. Temperature-dependent PL spectra revealed visible $X^0$ and T peaks, with a higher spectral weight of $X^0$ emission at elevated temperatures, implying reduced thermal stability of T in VP crystals. At lower temperatures, both $X^0$ and T emissions intensified, accompanied by blue-shifted peak positions, while the T/$X^0$ spectral weight ratio increased, indicating enhanced trion formation due to weakened phonon scattering and impurity-related processes. These results expand our understanding of VP's degradation behavior and pave the way for the development of robust and efficient optoelectronic devices leveraging VP's unique properties.

## Results and Discussion

### Excitation laser wavelength dependent photodegradation

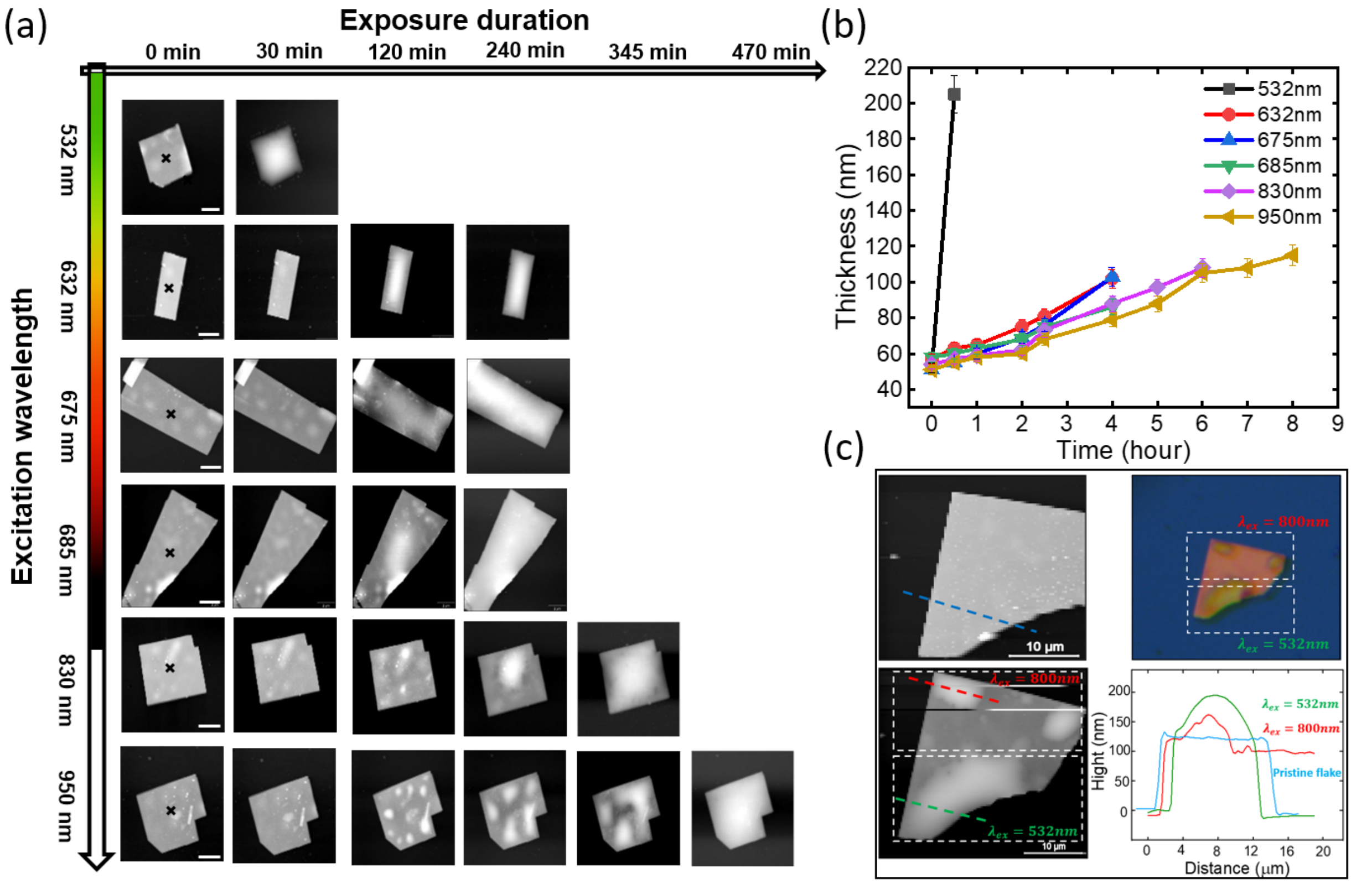

**Figure 1.** Effect of excitation wavelength on VP. a) AFM topographic maps of flakes overtime as a function of excitation wavelength. The degradation is observed by the appearance of bubble-like morphological features on the surface of the flakes. (b) Topographic height changes (thickness) taken from center of flakes which (marked by x on topography images) as function of time for different excitation laser wavelength. (c) Topographic map and optical image of a VP flake exposed to two different excitation wavelengths. (Top left) topography of the pristine VP flake. (Bottom left) topography of the flake after exposure to 532nm and 800nm. (Top right) shows the optical image of the flake after exposing to two different excitation lasers. The top side of the flake was exposed to 800nm light, and the bottom side of the same flake was exposed to 532nm light. (Bottom right) topographic line profiles of the different areas.

First, we investigate the effect of excitation laser wavelength on surface degradation of VP in the visible to near IR spectral range. To that end we prepared six VP (HQ Graphene) flakes of similar thicknesses between 50–55 nm by exfoliation on $SiO_2$/Si substrates. We then expose them to incident laser beam at different wavelength followed by AFM topographic measurements overtime (Fig. 1). For all the incident laser wavelengths used: 532, 632, 675, 685, 830, and 950nm, the power and time of exposure time were kept similar (~0.3 mW). Laser exposure of each flake was conducted in a dark room to rule out any influence of ambient light. AFM imaging of the flakes were undertaken immediately after exposure at every time step shown on the x-axis in Fig. 1. AFM topographic changes of VP as a function of excitation wavelength and exposure duration is shown Fig. 1a. A topographic increase in height is a mark of surface degradation as demonstrated by various previous works, where larger topographic change corresponds to more surface degradation[13, 21, 24-26]. The flake exposed to 532 nm incident laser degrades the fastest, oxidizing after 30 minutes (thickness changes as large as a factor of~4) Flakes exposed to 632nm, then 675nm, and finally, 685nm have a similar degradation topographic changes and duration time taken from center of flakes, marked by x on topography images (Fig. 1a&b). Bubble-like morphological features form on the surface initially and grow larger and after about ~4 hours the surfaces are bubble-like indicating extensive surface degradation. Surface bubble formation is a typical indicator of degradation in BP [25],[24],[27] which is similar to VP [21],[13]. Upon exposure to air, phosphorus present on the surface, edges, or defect sites of a material reacts with oxygen to form various phosphorus oxides which absorb moisture from the air, resulting in swollen, droplet-like protrusions on the material's surface. The flake exposed to an 830 nm excitation laser remained partially intact even after 6 hours, while the flake subjected to a 950 nm laser lasted for 8 hours in similar condition. In general, we note that the longer the laser wavelength (going from 532nm to 950nm) the slower the degradation time (see Fig. SI2 degradation time as a function of excitation wavelength). We also exposed a VP flake to two distinct excitation wavelengths consecutively first with 532 nm laser (energy above VP's bandgap) followed by 800 nm laser (energy below VP's bandgap). The topography of the flake before and after these two excitation wavelengths is depicted in Fig. 1c. The top left section of Fig. 1c illustrates the topography of the pristine VP flake, while the bottom left section showcases the topography of the flake after subjecting it to the two excitations. Initially, we exposed the top side of the VP flake to 800 nm light for an extended duration time, after which we imaged the flake's topography. Subsequently, we employed 532 nm light to illuminate the bottom side of the same flake for 30 minutes, followed by imaging the flake's topography. The optical image of the flake after exposure to these two distinct lights is presented in the top right section of Fig. 1c. Figure 1c (bottom right) exhibits the line profiles of different areas. These results clearly substantiate the effect of laser wavelength on degradation shown in Fig. 1a.

To further investigate photodegradation of VP, we tracked the Raman intensity of 12, 45, 11, and 35 nm thick flakes exfoliated onto $SiO_2$/Si substrate under continuous exposure to either a 514 nm (power ~220 µW) or 633 nm (power~ 4.5 mW) laser, which are above and near the bandgap of VP. At time equal zero all flakes show the typical VP Raman spectra, which contains numerous modes ranging from 100 to 300 $cm^{-1}$ and 350 to 500 $cm^{-1}$ originating from VP’s complex monoclinic crystal structure P2/n (13, $C_{2h}$(2/m))[23]. To compare all spectra, intensities were normalized by setting the Si substrate peak at ~520 $cm^{-1}$ to 1. Figure 2 a-d shows the evolution of Raman spectra with time for the 45 and 12 nm thick flakes exposed to the 514 nm laser and 35 and 11nm thick flakes exposed to the 633 nm laser. Figure 2e and f show how the normalized intensity of the two most intense VP modes, 354(S2 P9) and 359 (S1 P8) $cm^{-1}$, change with exposure to the 514 and 633 nm laser with time. Upon exposure to the 514 nm laser, the degradation process is quite rapid compared to the 633 nm consistent the wavelength dependent photo induced degradation observed using AFM.   Under exposure to the 514 nm laser, the spectra of the 45nm flake remains mostly unchanged for the first 90 sec. Afterward, the intensity decreases rapidly until t~350 sec when it reaches ~ 2% of the starting intensity. The intensity continues to decrease slowly becoming vanishingly low by 608 sec. In the 12 nm flake, the intensity immediately decreases reaching < 2% of the starting intensity in < 114 sec followed by a continuous slow decrease. As a comparison, the 11 and 35 nm flakes, which are exposed to the 633 nm laser at much higher power, show a significantly slower decrease in intensity. In the 11 nm thick film it takes more than 280 sec of exposure for the intensity to reduce to < 2% of the initial, while the intensity of the 35 nm flake reduces to 30 % after 800 sec. As the flakes degrade no new Raman peaks appear and only VP-related peaks are observed with reducing intensity. The expected surface phosphorus oxide species ($P_xO_y$) which have previously been observed [21] are not known to be Raman active, thus we would not expect the emergence of new peaks similar to degradation studies on black phosphorus[28].

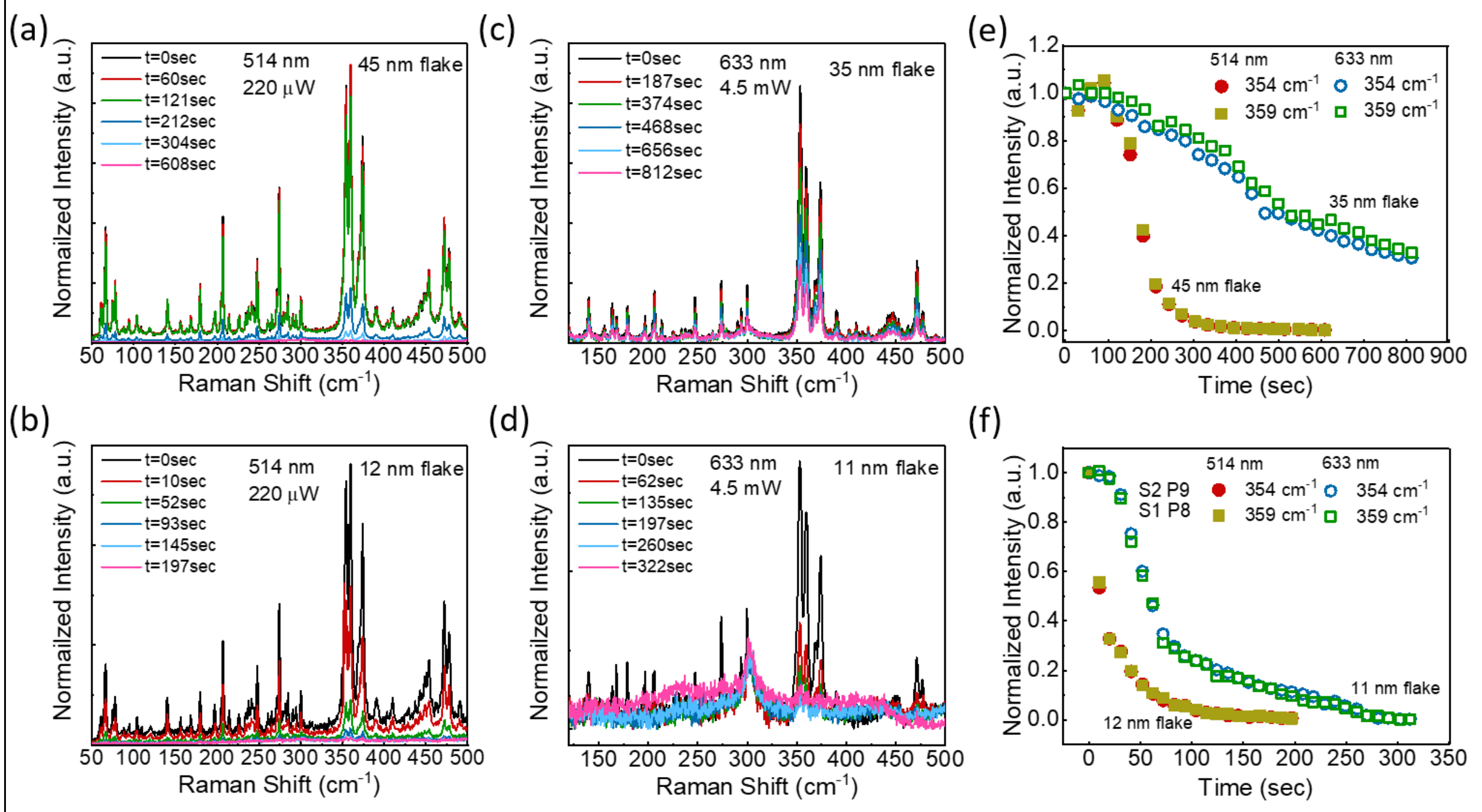

**Figure 2.** Laser accelerated degradation process. Raman spectra from a a) 45 nm and b) 12 nm VP flakes on $SiO_2$/Si substrate exposed to 220 μW 514 nm laser and c) 35nm and d) 11nm VP flake on SiO2/Si substrate exposed to 4.5mW 633nm laser over time. Spectra was normalized be setting the Si peak to 1. Evolution of Raman intensity of the 354 and 395 $cm^{-1}$ peaks over time for under exposure to 514 or 633 nm lasers e) 45 and 35 nm and f) 12 and 11 nm flake. The intensity was normalized to t=0 sec. peak intensity.

Based on excitation wavelength dependence studies in Fig. 1 and Raman measurements in Fig. 2, we conclude that when VP is exposed to light of energy greater than its bandgap, excited electrons and holes can create reactive species such as free radicals [29],[30], which can oxidize VP enhancing its degradation. Ahmed et. al[25] have shown that UV light is predominantly responsible for degradation of BP compared to longer excitation wavelengths. It was also proposed excitation wavelengths play drastically different roles in ROS generation, consequently affecting BP degradation. Similarly we find excitation wavelength dependent degradation of VP, it is plausible that ROS generation could also play a role in VP to undergo similar degradation mechanisms as BP under different excitation wavelengths[30]. We note however, as reported before [21], VP degrades slower compared to BP (see Fig. SI3 for comparison under similar experimental conditions). Our experiments show the extent of degradation depends on the excitation wavelength used and the duration of the excitation. The higher the energy of the photons, the more likely they are to generate reactive species that can cause degradation. When an excitation laser wavelength of 532 nm or shorter is used, degradation of VP is the fastest due to higher energy photons which can photo generate carriers in the VP likely to react with oxygen molecules in the air that can lead to oxidized phosphorus species. However, when the excitation energy is less than the VP's bandgap, it is less likely to generate ROS or other reactive radicals that can interact with VP. As a result, using longer wavelengths excitation (lower energy), does not degrade VP as fast.

**Excitons in VP**

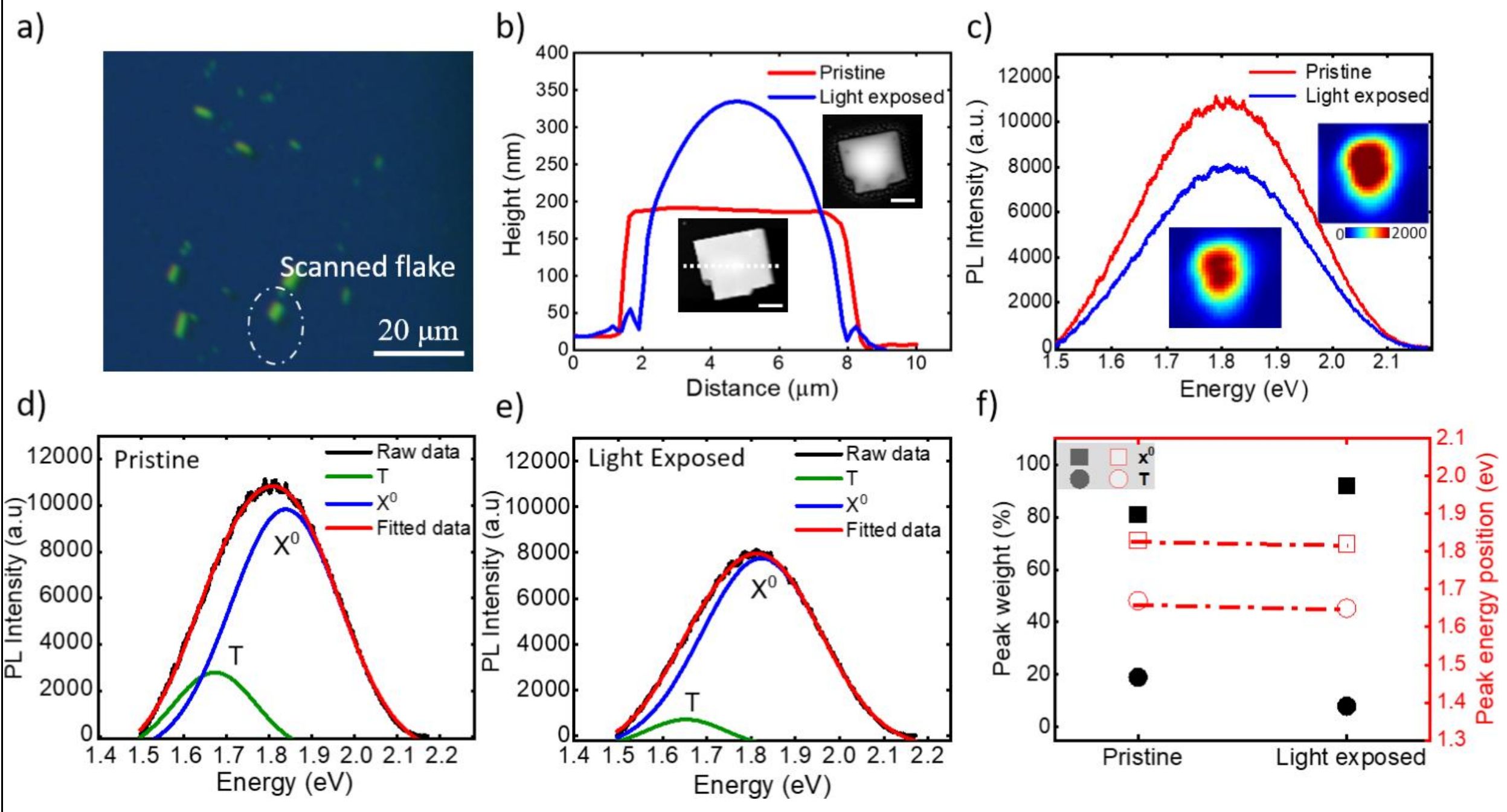

**Figure 3.** Topography and PL spectra overtime at room temperature. a) Optical far-field image of the VP flakes on $SiO_2$ substrate. b) Topography and line-profile images of the pristine and light exposed flake. c) PL spectra of pristine and light exposed VP flake. Images show corresponding hyperspectral PL images for pristine and light exposed flake. d) Fitted data of PL spectra for pristine VP and e) light exposed VP. f) Peak weight (black) and peak position energy (red) of the $X^0$ (square) and T (circle) emissions as a function of the time for the VP for pristine flake and light exposed. The wavelength of the excitation laser is 532nm. Scale bars in b denote 5 μm.

Degradation can shorten the lifetime of exciton in VP due to accelerated recombination at newly formed nonradiative centers, and impact their formation and stability, either promoting or inhibiting exciton creation. To understand the effect of degradation on changes in excitons we performed PL spectroscopy as a function of excitation power and sample temperature over an extended time. Fig. 3b shows topographic change of 160 nm exfoliated VP on $SiO_2$/Si substrate (flake indicated by a circle in the optical image in Fig. 3a) after exposure to a 532 nm laser beam for 30 minutes, the flake degrades substantially, and the thickness increases to 335 nm. Topographic height increase is commonly observed during degradation of van der Waals materials[25],[17]. In Fig. 3c the PL spectra shows that after light exposure, the intensity counts decrease compared to the pristine. The insets show 2D spatial maps extracted from hyperspectral PL data cubes (see Fig. SI 1). We note that the hyperspectral map resolution is diffraction limited however it still reproduces the topographic shape and reveals a homogeneous intensity change due to light exposure compared to the pristine flake. To further analyze the excitonic species of VP, we fit PL spectra taken at the center of the flake using dual-Gaussian function for the pristine and light-exposed flakes as shown in Fig. 3 d&e respectively. For the Pristine flake, the prominent peak (blue) with the energy bandgap 1.81 eV is assigned as the neutral exciton ($X^0$) emission, while the left shoulder (green) with an energy bandgap 1.67 eV is ascribed to the charged exciton or trion (T) emission as reported before[17]. After light exposure, the exciton peak appears at 1.82eV, and the trion peak at 1.68 eV. Figure 3f (right panel), shows the peak energy position for both $X^0$ and T for the pristine VP flake and light exposed flake. The changes in peak position for both $X^0$ and T are negligible due to uncertainties of the spectral fitting. However, As the exposure time increases, the trion emission gradually decreases compared to the initial state. Based on the fitted curve areas we calculated the peak weights [17],[31] which is the ratio of the trion ($\frac{T}{T+X^0}$) and ratio of the neutral-exciton ($\frac{X^0}{T+X^0}$) for both pristine VP flake, and light exposed flake for an extended time. We found that the PL spectral weights of the $X^0$ and T emissions gradually changes from $X^0$: 80.5%, T: 20.5% in the pristine sample to $X^0$: 92%, T: 8% for the degraded sample as shown in Fig. 3f (left panel), suggesting a gradual decline of the trion population. Degradation lessens exciton lifetime through faster recombination and alter their formation and stability, impacting quantum efficiency.

**Power dependence for PL spectra of VP**

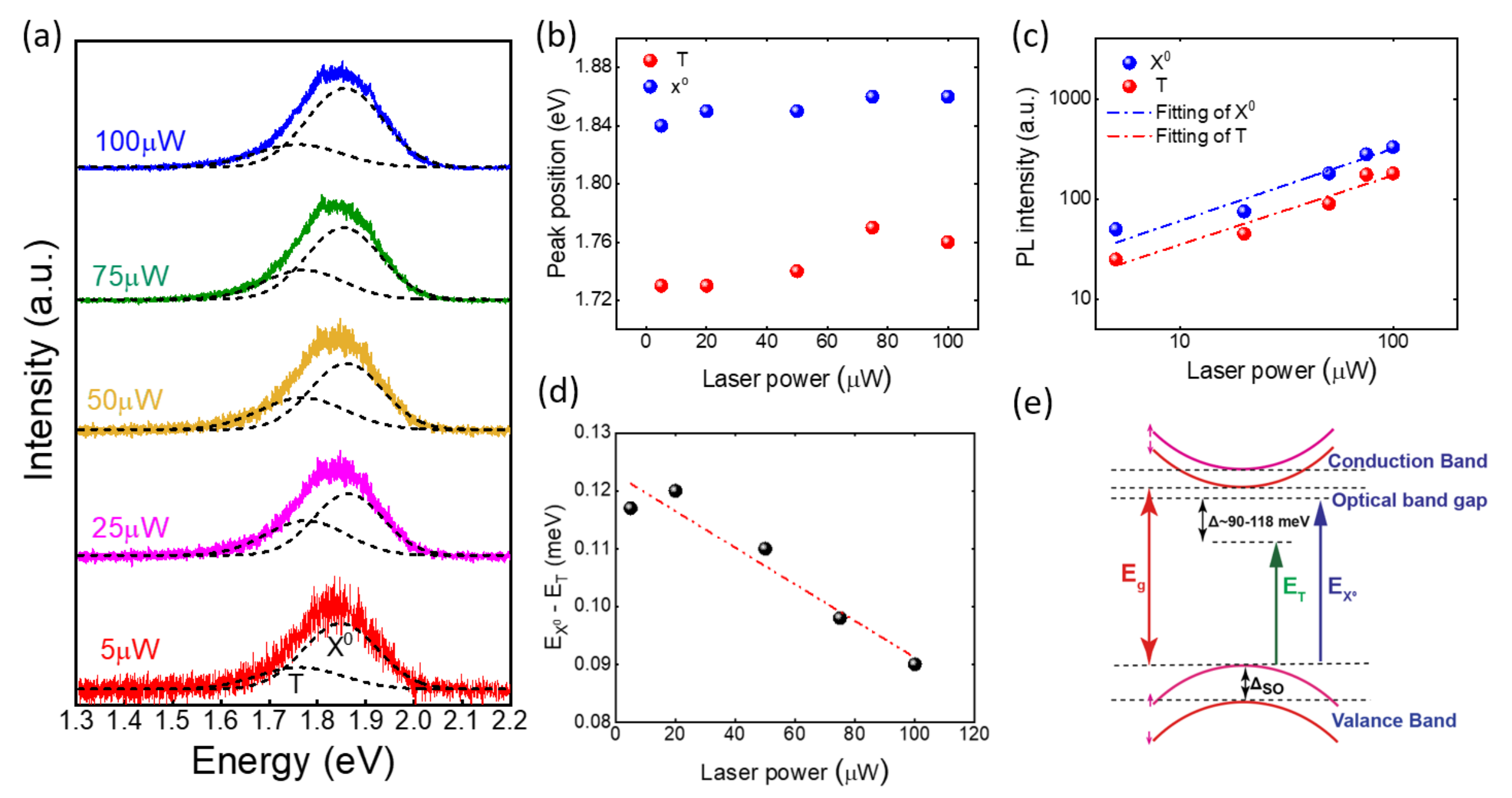

**Figure 4.** Laser power dependent $X^0$ and T emissions at low temperature (T=4 K). (a) PL spectra at different powers. The black dashed lines are based on the dual-Gaussian fitting. (b) Peak positions at different powers. (c) Integrated PL intensities for two peaks at different powers, dashed lines show linear fits for two peaks. (d) The difference in the PL peak energies of excitons ($X^0$) and trions (T) in (a) , $E_{X^0} - E_T$ (symbols), as a function of excitation power. The red dashed line is a linear fit to the laser power-dependence. (e) Schematics of the energy diagram for neutral exciton and trion of VP.

The evolution of the PL spectra as a function of the excitation power gives information concerning the nature of the observed emissions. To that end we have measured the power-dependent evolution of the PL spectra of a VP flake at T = 4 K. A 532 nm laser was used as the excitation incident light with different laser power shown in Fig. 4. We used Dual-Gaussian fittings to the integrated PL spectra and extracted the spectral components of these two peaks, indicated by black dash lines in Fig. 4a and in Fig. 4b. We show the PL peak energy position of the $X^0$ and T emissions as a function of the laser power. When the excitation power changes from 5μW to 100μW, the optical gap for neutral-exciton increases ~ 15meV, and for trion it increases ~ 33meV. As the density of the photoexcited carriers increases with increasing excitation power, the quasi-Fermi levels for electrons and holes shift respectively into the conduction and valence bands leading to a blue shift in other 2D materials such as TMDs [32],[33, 34]. However, many-body interactions caused by heating or strain-induced band variation giving rise to reduced bandgap and exciton binding energy would lead to a redshift of the excitonic peaks [35],[36]. For the low excitation powers used in our experiments presumably the carrier density is not high enough to give rise to the many-body interactions and the observed blue shifts are essentially caused by the shift of the quasi-Fermi levels. Figure 4c shows the PL intensities at different powers for both neutral exciton and trion peaks of VP. Based on the fitted data at different exciton powers, both intensities increase approximately linear with the power, as demonstrated by the fits to the power law $I = P^k$, with k = 0.88 for neutral exciton and k = 0.79 for trion peak. Figure 4d shows the difference in the PL peak energies of excitons ($X^0$) and trions (T), $E_{X^0} – E_T$ (symbols), as a function of laser excitation power, and the red dashed line is a linear fit to the laser power-dependence in Fig. 4d shows PL intensities of the $X^0$ and T as a function of the laser power. Interestingly, the energy difference decreases from ~ 118 meV to ~90 meV when the power excitation laser changes from low to high power due to alteration of the energy levels of excitons introduced due to degradation at higher laser power[37]. We note to reduce degradation of VP, we used low power measurements (5-100μW). However, it is been reported[17] that the trion peak saturates at high laser powers (~2mW).

**Temperature Dependence**

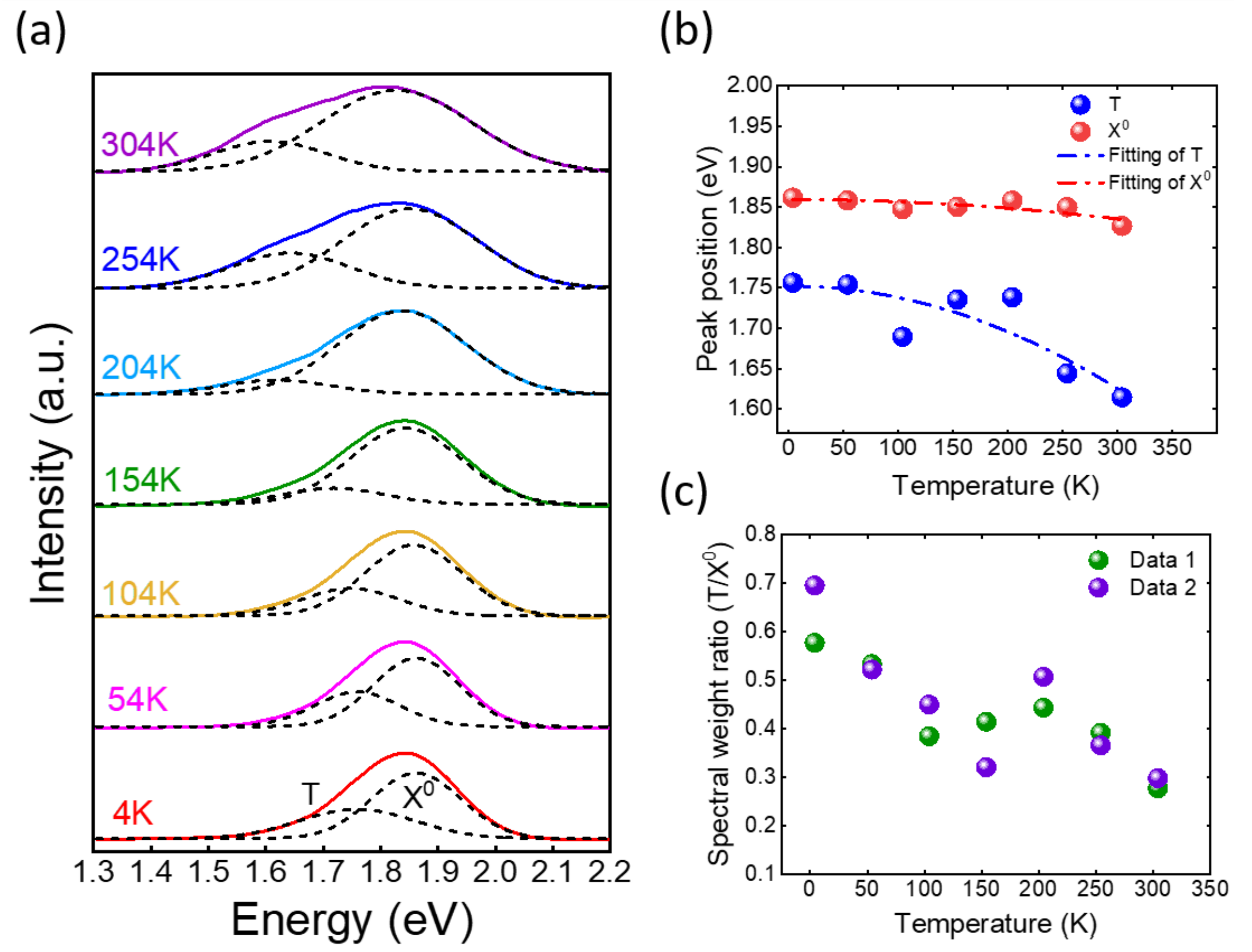

**Figure 5.** The extracted peak energy of $X^0$ and T emissions as a function of temperature. The dash dot lines are fittings. Evolution of $X^0$ and T emissions as a function of temperature and the calculation of trion binding energy. (a) Temperature-dependent evolution PL spectra of VP by representative deconvolution of the $X^0$ and T (black dash lines) emission components. (b) peak

position of $X^0$ and T at different temperatures. The dash lines are fitted data. (c) spectral weight ratio of two different flakes.

The temperature-dependent PL spectra can reveal insights into the tunability of the energy bandgap due to thermal changes and the stability as well as band assignments of the $X^0$ and T peak components. Figure 5a depicts the evolution of temperature-dependent of PL spectra when excited by a 532 nm laser. At temperature of 304 K, visible $X^0$ and T peak components may be seen, with a greater spectral weight of $X^0$ emission, indicating reduced thermal stability of T in VP crystals, which has been confirmed in BP and $MoS_2$ crystals [38],[39]. Both the $X^0$ and T emissions grow more intense when the temperature decrease to 4 K, which is attributed to less phonon scattering at low temperatures[40]. In addition, for both $X^0$ and T emissions, there are apparent blue-shifts of the peak locations. The peak emission energy of $X^0$ and T as obtained by dual Gaussian fitting is shown in Fig. 5b. The emission energy blueshifts are consistent with the O'Donnell equation, which predicts the temperature-dependent bandgap of semiconductors based on electron-phonon interaction[41],[42].

$$E_g(T) = E_g(0) - S < \hbar\omega > \left[\coth\left(\frac{<\hbar\omega>}{2kT}\right) - 1\right] \tag{1}$$

Where $E_g(0)$ is the ground-state transition energy at 0 K, S is a dimensionless coupling constant and $< \hbar\omega >$ is and average phonon's energy. According to the fitting lines of $X^0$ and T, we discovered that the $X^0$, and T emission energy is 1.86 eV, and (1.75 eV) at 4 K, respectively (fig. 5b). The Trion dissociation energy $< \Delta\hbar\omega >$ is 110 meV at 4K. The dissociation energy steadily increases to 260 meV at 304 K as the temperature rises which is consistent to previous reports [17].

Figure 5c shows the ratio of the spectral weight of trion to exciton ($T/X^0$) changes from 0.28 at 300 K to 0.69 at 4 K. The significant increase in the trion spectral weight at low temperatures, implies the formation of more trion, might be associated with the weakening of phonon scattering at these temperatures. This weakening could be due to suppressed lattice vibrations at lower temperatures, thereby facilitating the formation and stability of trion. Also, impurity related emission processes dominate the electron–hole pairs' recombination. With impurities in the VP flakes, negative T could form by capturing the doped electrons. As the temperature increases, thermal fluctuations can disturb the binding of the trion and reduce their spectral weight ratio. Thus, a strong T emission can be observed at low temperatures, which has also been seen in other 2D materials like TMDs[40],[43]. Based on our results for two different flakes, the maximum ratio at 4K was 0.70 and 0.59 which is comparable with other works[17].

Finally, we investigate the temperature dependent phonon properties of VP using temperature dependent Raman from 173 to 673 K. Two VP flakes exfoliated on $SiO_2$/Si with thickness of 23 and 53nm were characterized. Figure 6a show the evolution of the Raman spectra with temperature increasing from 173 to 673 and back down to 303 K on the 23nm thick flake. As the temperature increased a clear broadening and red shifting of peaks was observed with many of the peaks merging. At 673K a new broad peak between 300 and 350 $cm^{-1}$ begins to emerge, which become quite pronounced upon cool back to 303K. We notice that not all the peaks in the original spectra can be observed on cooling and the intensity of the observable peaks is much lower indicating degradation of the sample. Most striking is the emergence of two new broad peaks at ~210 and 330 $cm^{-1}$. At this point the origin of these peaks remains unknown and cannot be assigned to any of the known phosphorus allotropes.

The temperature dependent shift for 9 VP peaks from the 23 and 53 nm thick flakes are plotted in Fig. 6b. All modes were found to have a linear temperature dependence and were fit to $\omega(T) = \omega(0) + \chi T$, which describes the first order temperature coefficient $\chi$. The peaks position vs. temperature for both flakes is consistent across the measured temperature range with $\chi$ ranging from 0.0083 to 0.0277 $cm^{-1}$/deg. These $\chi$ values are consistent with those reported by L. Zhang et al.[23] except for the $T_g$ and P tube modes, which L Zhang et al. found to have a non-linear temperature dependence. However, we found the $T_g$ and $P_{tub}$ modes merged above 473 K making it challenging to accurately determine peak position. Upon cooling back to 303 K the peak positions were found to be only slightly lower than when measured during heating indicating minimal structural degradation.

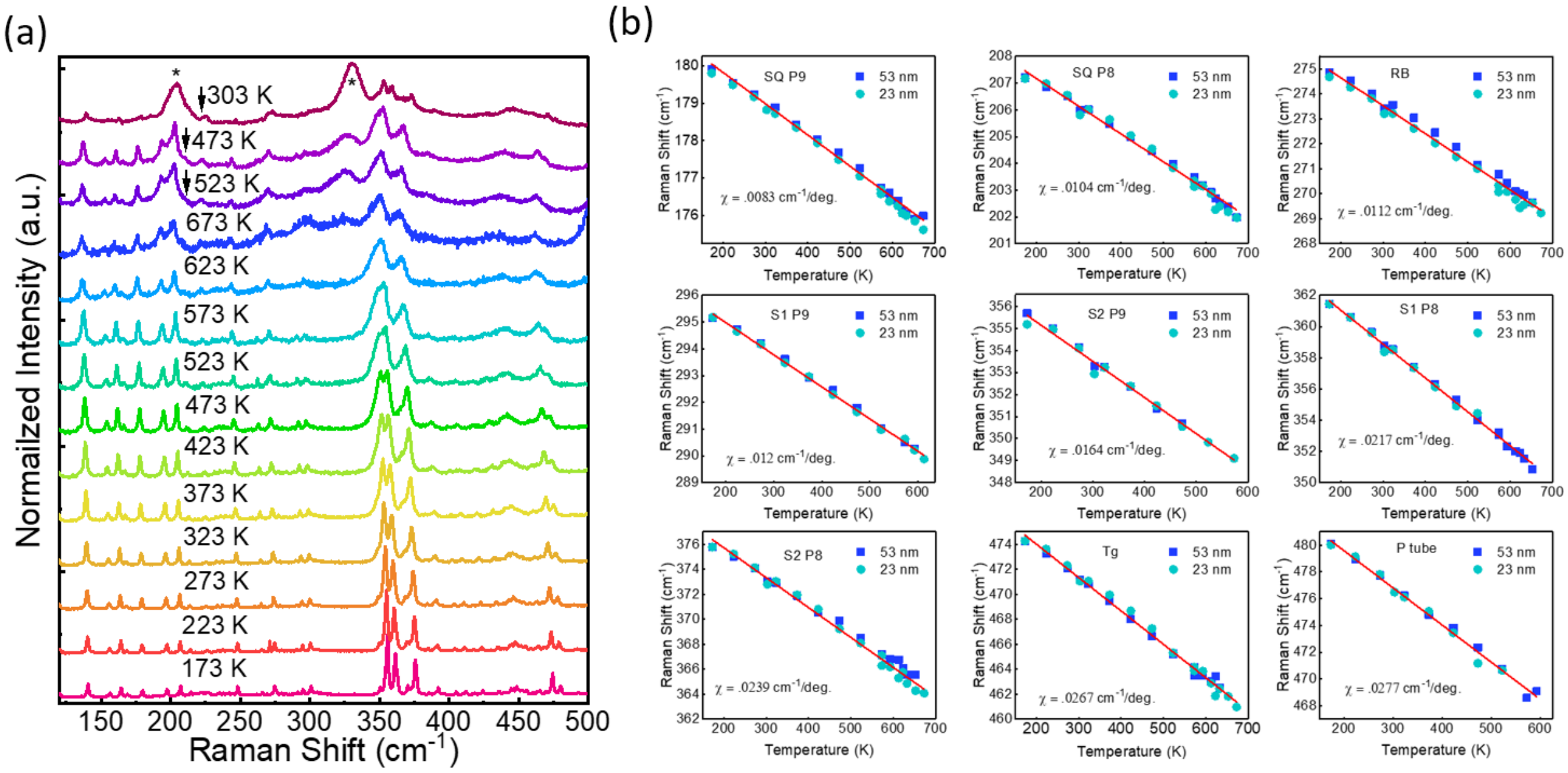

**Figure 6.** (a) Normalized Raman spectra of violet phosphorus single crystals in a temperature range of 173−673 K and (b) The temperature-dependent shift for 9 VP peaks from the 23 and 53 nm thick flakes are plotted with the corresponding fitting lines.

## Conclusion

We report on a thorough investigation of the photo- and thermal-degradation of VP, revealing that high-energy laser exposure (above bandgap, >532-nm) significantly accelerates degradation. PL spectroscopy was performed as a function of excitation energy, power, and temperature to understand the effect of degradation on excitons. Degradation can shorten the lifetime of excitons, and impacts exciton formation and stability. Temperature dependent Raman was used to investigate the phonon properties of VP. VP flakes were shown to survive up to 675 K with some irreversible degradation. Our findings elucidate the primary factors causing VP degradation, offering insights that will support the development of durable, long-lasting optoelectronic devices based on VP.

## METHODS

### Optical Measurements

**Raman Measurements.** Raman spectra was collected using Renishaw Inviva system 20µm slits and 3000 line/mm grating. For temperature dependent measurements a Linkam environmental stage was used to control temperature under an $N_2$ environment and 220µw 633nm excitation source was used. Samples were cooled to 173 K under flowing $N_2$, heated to 673 K and cooled back to 303 K taking measurements both during heating and cooling. Measurements between 173 and 573 K were taken at 50 ° increments and at 20 ° increments from 573 to 673 K. Before each measurement the temperature was stabilized for 10mins. For photodegradation measurements VP flakes were continuously exposed to either a 514 nm (power ~220 µW) or 633 nm (power~ 4.5 mW) laser under an ambient atmosphere and Raman data collected periodically.

**Photoluminescence (PL) Measurements**. PL spectra were measured using confocal laser scanning microscope system equipped with a vibration-free closed-cycle cryostat (Attodry 800, attocube). CW laser as an excitation source was focused into a small spot with a diameter of approximately 2-3 μm on the sample through a 100× objective lens (APO/VIS, N.A. = 0.82; attocube) inside the vacuum chamber. The PL spectra was then collected by the same lens and filtered the excitation beam using long-pass filter before entering a spectrometer (Andor) which consisted of a monochromator and a thermoelectrically cooled CCD camera. The representative violet phosphorus flakes are exposed to discrete wavelengths (532, 633, 565, 675, 685, 830 and 950 nm) using commercial Matisse single frequency tunable laser pumped by 532nm CW laser (Spectra-Physics.co.)

**Author Contributions**

Y.A. conceived and guided the experiments. M.G., S.S., S.G., and Y.A. carried out AFM, PL measurements at room and low temperature. T.P. and M.S. carried out Raman measurements. M.G. analyzed experimental data of PL measurements. M.S. analyzed Raman measurements.

**FUNDING**

M.G., S.S., S.G. and Y.A. acknowledge support for this work is provided by the Air Force Office of Scientific Research (AFOSR) grant number FA9550-19-0252. M.S. and T.P. acknowledge funding from the Air Force Office of Scientific Research (AFOSR) under award number FA9550-19RYCO.